\documentclass[aps,floatfix,eqsecnum,showpacs,preprint,tightenlines,nofootinbib]{revtex4-1}
\usepackage{epsfig}

\usepackage{amsmath}
\usepackage{amssymb}
\usepackage{color}

\begin{document}



\title{Applications of the chiral potential with the semi-local regularization in momentum space
to the disintegration processes}
\author{V.~Urbanevych}
\affiliation{M. Smoluchowski Institute of Physics, Jagiellonian University, PL-30348 Krak\'ow, Poland}
\author{R.~Skibi{\'n}ski}
\affiliation{M. Smoluchowski Institute of Physics, Jagiellonian University, PL-30348 Krak\'ow, Poland}
\author{H.~Wita{\l}a}
\affiliation{M. Smoluchowski Institute of Physics, Jagiellonian University, PL-30348 Krak\'ow, Poland}
\author{J.~Golak}
\affiliation{M. Smoluchowski Institute of Physics, Jagiellonian University, PL-30348 Krak\'ow, Poland}
\author{K.~Topolnicki}
\affiliation{M. Smoluchowski Institute of Physics, Jagiellonian University, PL-30348 Krak\'ow, Poland}
\author{A.~Grassi}
\affiliation{M. Smoluchowski Institute of Physics, Jagiellonian University, PL-30348 Krak\'ow, Poland}
\author{E.~Epelbaum} 
\affiliation{Ruhr-Universit\"at Bochum, Fakult\"at f\"ur Physik und Astronomie, Institut f\"ur Theoretische Physik II, D-44780 Bochum, Germany}
\author{H. Krebs}
\affiliation{Ruhr-Universit\"at Bochum, Fakult\"at f\"ur Physik und Astronomie, Institut f\"ur Theoretische Physik II, D-44780 Bochum, Germany}

\date{\today}

\begin{abstract}
 We apply the chiral potential with the momentum space semi-local regularization 
 to the
 $^2$H and $^3$He photodisintegration processes and to the (anti)neutrino induced
 deuteron breakup reactions. Specifically, the differential cross section, the photon analyzing power 
 and the final proton polarization have been calculated for the 
 deuteron photodisintegration at the photon energies 30 MeV and 100 MeV. 
For the $^3$He photodisintegration predictions for the semi-inclusive and
 exclusive differential cross sections are presented for the photon energies up to 120 MeV. 
The total cross section is calculated for the (anti)neutrino disintegrations of the deuteron
for the (anti)neutrino energies below 200 MeV. 
The predictions based on the Argonne V18 potential 
or on the older chiral force with regularization applied in
coordinate space are used for comparison. 
Using the fifth order chiral nucleon-nucleon potential 
supplemented with dominant contributions from the sixth order allows us to
obtain converged predictions for the regarded reactions and observables.
Our results based on the newest semi-local chiral potentials show even smaller cutoff dependence 
for the considered electroweak observables than the previously reported ones with a
coordinate-space regulator.
However, some of the studied polarization observables in the deuteron photodisintegration process reveal
more sensitivity to the regulator value than the unpolarized cross section. 
The chiral potential regularized semi-locally in momentum space yields also fast convergence of results with the chiral order. 
These features make the used potential a high quality tool to study electroweak processes.
\end{abstract}


\maketitle


\section{Introduction}

Chiral effective field theory ($\chi$EFT)
is nowadays the most reliable approach to study low-energy nuclear forces.
It's continuous development for nearly 30 years resulted 
in advanced two-nucleon and 
  many-nucleon interactions~\cite{reinkrebs2018,EpelHam2008, Machleidt2011, Piarulli2012, 
  Epelbaum2014SCS, Bernard2008, Bernard2011}.

In 2018 the Bochum group presented a new version of the chiral interaction \cite{reinkrebs2018}, 
which differs from the previous realization, among others, by
the regularization scheme applied directly in momentum space. 
This semi-local momentum-space (SMS) regularized potential
has been derived completely
up to the fifth order (N$^4$LO) of the chiral expansion and
some contributions from the sixth order have been even included in its ``N$^4$LO$^+$'' version.
Up to now this potential has been used out of necessity
in the two-nucleon (2N) system to fix its free parameters 
and later applied to nucleon-deuteron elastic scattering as well as to
the nucleon induced deuteron breakup process~\cite{Epelbaum2020,
Volkotrub2020},
delivering at the N$^2$LO a data description of the similar quality as the non-chiral semi-phenomenological potentials.
In this paper we extend applications of the SMS interaction beyond the purely strong processes to a few electroweak reactions. 

Investigations of the electroweak processes with the older chiral force with the non-local regularization resulted
  in predictions strongly dependent on the cut-off parameter 
  (see e.g. \cite{Skib2006,Skib2014,Rozpedzik2011}). 
  Such a picture was observed with the single-nucleon current and with inclusion of some meson-exchange currents.
  In 2016 we showed~\cite{skibgol2016} that semi-local regularization (applied in coordinate space) 
helps to avoid such big variations with respect to
  the regulator, but predictions were still clearly dependent on the cutoff parameter.
  In the present paper we check if properties of the SMS potential match or possibly even surpass those of the former version of the chiral potential.
  To this end
  we apply the SMS chiral potential to the selected  electromagnetic and weak processes. 
Specifically we study the $\gamma + ^2{\rm H} \rightarrow {\rm p+n}$,
 $\nu_e + ^2$H~$\rightarrow \nu_e + {\rm p+n}$,
 $\bar{\nu}_e + ^2$H~$\rightarrow \bar{\nu}_e + {\rm p+n}$,
 $\bar{\nu}_e + ^2$H~$\rightarrow e^+ + {\rm n+n}$ and
 $\gamma + ^3{\rm He} \rightarrow {\rm p+p+n}$ reactions.
  For the sake of comparison we also use the results of
  our previous research with the 
semi-local coordinate-space (SCS) regularized potential
\cite{Epelbaum2014SCS,Epelbaum2014EPJA} 
 and with the Argonne V18 (AV18) potential~\cite{AV18Wiringa}.

  One of the most challenging problems with the application of a nucleon-nucleon (NN) potential 
to electroweak processes with nuclear systems is a construction of consistent two-nucleon (2N)
and, more generally, many-nucleon electroweak current operators~\cite{Epelbaum2020Front}.

Many-nucleon currents linked to various models of nuclear forces
have been investigated for a long time 
(see e.g.~\cite{RISKA1972, Riska1985, Carlson1997, Arenhovel2001, Marcucci2005, Kolling2011,BaroniSchiavilla2017}).
  No full 2N electromagnetic or weak current operator consistent with the SMS chiral force has been derived yet,
see however~\cite{FilinPhysRevLett} for the recent calculation of the deuteron structure radius with the 
consistent 2N charge operator. 
Thus we rely on the single-nucleon current (SNC) and use the Siegert theorem 
to take into account many-body contributions to the nuclear electromagnetic current. 
Many-nucleon currents, even included implicitly, 
are absolutely indispensable in a correct treatment of photonuclear reactions.
It has been shown (see for example Ref.~\cite{Golak2005}) that
their omission leads to incorrect predictions, especially for polarization 
observables.
In the case of weak reactions 
we use the non-relativistic form of the single-nucleon weak current operator, whose
components are defined in Ref.~\cite{Golak2014}. The dominant role of the single-nucleon weak current in neutrino scattering
of the deuteron has been recently demonstrated by Baroni and Schiavilla~\cite{BaroniSchiavilla2017} 
who have found only a few percent
contribution from higher order current operators. 
 
The rest of this paper is structured as follows. 
  In the next section we briefly describe our formalism and then in Sec.~III we show
our results for the deuteron photodisintegration and (anti)neutrino induced deuteron 
disintegration processes. Section~IV comprises our predictions for $^3$He disintegration. 
We summarize and conclude in Sec.~V.

\section{Theoretical formalism}

 Our approach, which is based on the Schr\"odinger and Lippmann-Schwinger equations (for 2N reactions) and on
 the Faddeev formalism (for 3N reactions) has been described in detail in  
 \cite{GolakKamad2000_ExplDescr, Golak2014, Golak2005}. 
 In short, the path to the observables for the 
 electromagnetic or weak disintegrations leads through the appropriate nuclear matrix elements.
 
 For the deuteron photodisintegration process, in the nuclear matrix elements 
 \begin{equation}
  N^\mu_{deu} \equiv \langle \Psi^{2N}_{scatt} \mid j^\mu_{2N} \mid \Psi^{2N}_{bound} \rangle,
  \label{NmuDeu}
 \end{equation}
a full 2N electromagnetic current operator
appears between the initial
 deuteron bound state $\mid \Psi^{2N}_{bound} \rangle$ 
and the final 2N scattering state $\mid \Psi^{2N}_{scatt} \rangle$.
 In order to obtain the deuteron bound state we solve the
 Schr{\"{o}}dinger equation with a given 2N potential $V$. Further, the scattering state is
constructed from a solution of the Lippmann-Schwinger equation for the $t$ operator:
$t = V + tG_0V$, where $G_0$ is a free 2N propagator. Using this equation~(\ref{NmuDeu}) takes the form:

 \begin{equation}
  N^\mu_{deu} = \langle \vec{p_0} \mid (1 + t G_0) j^\mu _{2N} \mid \Psi^{2N}_{bound} \rangle,
  \label{NmuDeu2}
 \end{equation}
 where $\mid \vec{p_0} \rangle$ is the antisymmetrized eigenstate of the relative proton-neutron momentum.  

Our formalism for the (anti)neutrino induced deuteron disintegrations is essentially the same~\cite{Golak2014,Golak2018,Golak2019}. 
Especially, for the neutral-current (NC) driven processes the isospin structure of the current operator and the 2N final state
are the same as for the photodisintegration reaction. In the case of the charged-current (CC) driven reaction
only some straightforward modifications are introduced in the corresponding weak single-nucleon current operator. 

 For the $^3$He photodisintegration with three free nucleons in the final state, if the 3N force is neglected,
 the nuclear matrix elements are given as \cite{Golak2005}:

 \begin{equation}
  N^\mu_{3N} = \langle \Phi_{3N} | (1 + P) j^\mu _{3N} | \Psi^{3N}_{bound} \rangle + \langle \Phi_{3N} | (1 + P) | U^\mu \rangle \, ,
  \label{Nmu3N}
 \end{equation}
where 
$ \mid  \Psi^{3N}_{bound} \rangle $ represents the initial 3N bound state and
$\mid \Phi _{3N} \rangle$ is an antisymmetrized state which describes the free motion of three outgoing nucleons.
Further, the permutation operator $P$ is built from transpositions $P_{ij}$ of particles $i$ and $j$ : 
$P = P_{12}P_{23} + P_{13}P_{23}$ and the auxiliary state $\mid U^\mu \rangle$ allows us to include all 
the final state interactions among the three outgoing nucleons.
It is a solution of the Faddeev-like equation \cite{Golak2005}
which reads

\begin{equation}
 | U^\mu \rangle = t \tilde{G_0}(1+P)j^\mu_{3N} | \Psi^{3N}_{bound} \rangle + t\tilde{G_0}P|U^\mu \rangle,
 \label{Fadeev}
\end{equation}
where $\tilde{G_0}$ is a free 3N propagator and $j^\mu_{3N}$ is the total 3N electromagnetic current operator. 

At the moment 2N currents, with the same regularization as used in
the SMS interaction investigated here, are not available. 
While their operator form has been already derived~\cite{Krebs2019,Kolling2009PhysRevC,Kolling2011,Krebs2016,Krebs2019FewBodySyst,Krebs2020},
a consistent regularization of these currents 
is still under development.  
Thus for the weak processes we use contributions from the single-nucleon currents only. 
In the case of the electromagnetic reactions 
additional contributions are taken implicitly into account using the Siegert theorem, as described in detail in Ref.  
\cite{GolakKamad2000_ExplDescr,Golak2005,Rozpedzik2011}. We partly substitute 
electric multipoles by the Coulomb ones, calculated from the single-nucleon charge
density operator. To this end we carry out the multipole decomposition of the 
corresponding SNC matrix elements.

We perform our calculations in the momentum space, in the partial wave decomposition scheme.
For the deuteron disintegration we take into account all partial waves in the 2N system up to the total angular momentum $j= 4$.
For $^3$He photodisintegrations we use all two-nucleon partial waves up to the total 2N angular momentum $j= 3$ and 
all three-nucleon partial waves up to the total 3N angular momentum $J=\frac{15}{2}$.
For further details on our computational scheme see Ref.~\cite{Golak2005}.

\section{Results for $^2$H disintegration}

\begin{figure}[!htb]%
  \begin{center}
  \includegraphics[width=1\textwidth,clip=true]{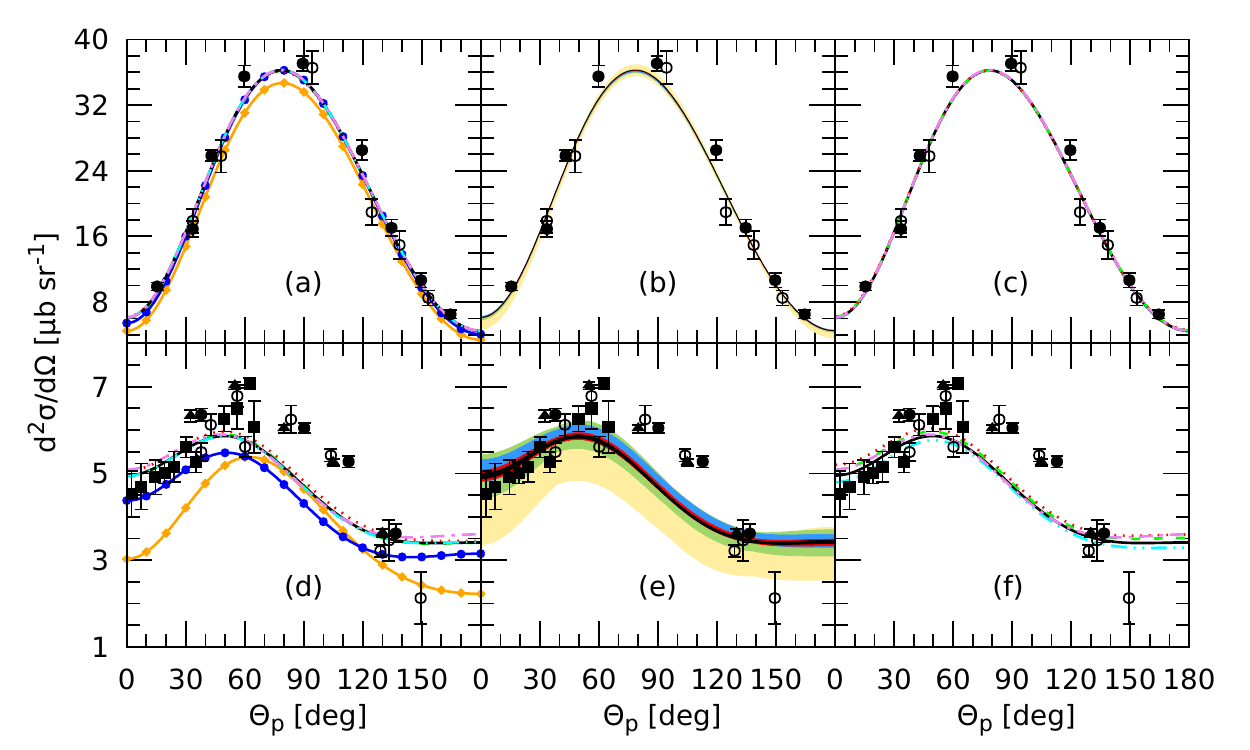}
  \end{center}
  \caption{The differential cross section $\frac{d^2\sigma}{d\Omega}$ for the
  $\gamma$~+~$^2$H~$\rightarrow$~p~+~n reaction at the laboratory photon 
  energy E$_\gamma$~=~30~MeV (top)
  and E$_\gamma$~=~100~MeV (bottom) as a function of the proton emission angle 
	$\theta_p$ (the angle between the initial photon momentum and the 
	proton momentum) 
   in the center of mass frame.
  The left column shows the dependence of predictions on the order of chiral expansion.
  The orange solid line with diamonds, the blue solid line with circles, the green dashed, the red dotted, the black solid,
  the cyan double-dot-dashed and the violet dot-dashed curves
  correspond to the LO, NLO, N$^2$LO, N$^3$LO, N$^4$LO, N$^4$LO+ and AV18 potential based
  predictions, respectively.
  Truncation errors for the different orders of chiral expansion are presented in the middle column.
  The yellow band shows the truncation errors at the NLO, green - at the N$^2$LO, blue - at the N$^3$LO, red - at the N$^4$LO and black
  one at the N$^4$LO$^+$ orders. The right column shows the chiral SMS predictions at N$^4$LO,
  calculated using different values of the cut-off parameter $\Lambda$.
  The cyan double-dot-dashed, the black solid, the green dashed, the red dotted and
  the violet dot-dashed curves correspond
  to $\Lambda$ = 400~MeV, 450~MeV, 500~MeV, 550~MeV, and to the AV18 based predictions respectively.
  For the predictions shown in the left and in the middle columns the regulator value $\Lambda=450$~MeV
is used.
  The data points at 30~MeV are the same as in \cite{Ying_Experiment_Deut} and at 100~MeV are 
taken from \cite{Ying_Experiment_Deut}
  (open and filled circles, squares) and from \cite{DeSanctis_Experiment_Deut}(triangles).}
  \label{cross}%
\end{figure}

First we discuss our results for the 
deuteron photodisintegration process $\gamma$~+~$^2$H~$\rightarrow$~p~+~n 
at two laboratory photon energies E$_\gamma$ = 30~MeV and E$_\gamma$ = 100~MeV. 
 Figure~\ref{cross} shows the differential cross section 
 $\frac{d^2\sigma}{d\Omega}$ obtained using the chiral SMS potential for both energies.
 For the sake of comparison we also
 show predictions obtained using the AV18 NN potential \cite{AV18Wiringa}.
 Both for the chiral SMS force and for the AV18 potential our predictions were obtained
with the single-nucleon current supplemented implicitly by some 2N parts,
 using the Siegert theorem \cite{Skib2006}.

 In the left column of Fig.\ref{cross} we present results calculated at
 different chiral orders (from LO to N$^4$LO$^+$) with the regularization parameter $\Lambda$~=~450~MeV.
 For the energy E$_\gamma$~=~30~MeV (top row) only the LO prediction is
 noticeably separated from all others and the difference among remaining predictions is very small
 ($\approx 0.06$\% at the maximum of the cross section  
between N$^2$LO and N$^3$LO results) and even less 
for all subsequent chiral orders.
 It shows that for this photon energy the SMS potential based predictions converge rapidly 
 and contributions from high orders are not crucial.
 At 100~MeV our predictions converge more slowly, but starting from N$^2$LO all the curves are
 very close to each other (the difference between the lines remains below $3\%$).
 In the case of E$_\gamma$ = 100~MeV the data description is worse than at E$_\gamma$ = 30~MeV, but
 based on the semi-phenomenological results by Arenh\"ovel {\em et al.}~\cite{ArenhovelPhotodisint1991}
it is expected that for higher energies 2N electromagnetic currents contribute substantially and we thus expect that 
 our predictions will improve significantly when explicit 2N current operators, fully consistent
with the 2N potential, are included.

 The middle column of Fig.\ref{cross} presents the truncation errors arising, at a given chiral order, due to 
 neglecting of higher-order contributions to the chiral potential. These theoretical estimates
 were calculated employing the prescription advocated in \cite{Binder2015} and later used also
 for electromagnetic reactions with the SCS potential in~\cite{skibgol2016}
\footnote{It is also possible to perform a more sophisticated estimation based
on the Bayesian approach~\cite{Furnstahl2015, Epelbaum2020}, however, for
the sake of comparison with~\cite{skibgol2016} we use the
prescription~\cite{Binder2015} in this work.}.  
 The observed picture demonstrates that only tiny contributions from
 the potential components above N$^4$LO should be expected, as the band showing the truncation error for the highest
 presented chiral order N$^4$LO$^+$ is quite narrow - its width in the maximum of the cross section is around 0.1$\%$ (2.5\%)
of the cross section magnitude at the  photon energy $E_\gamma$ = 30~MeV (100~MeV). Notice, however, that the estimates
of truncation error may change upon performing a more complete treatment of the current operators.

 Finally, the right column of Fig.\ref{cross} shows the dependence of predictions on the value of the
 regularization parameter $\Lambda$ in the range $\Lambda \in [ 400, 550 ] $~MeV. 
 It is clearly seen that the SMS potential
 provides us with a weak dependence of the predicted cross section on the cut-off parameter at both investigated energies.
 This is an important improvement compared to
 the older versions of the chiral force (see e.g. \cite{skibgol2016}), where regularization strongly influenced the results, 
 yielding a spread of predictions up to 20\% for small proton scattering angles at $E_\gamma = 100$~MeV. 

\begin{figure}[h]%
  \begin{center}
  \includegraphics[width=0.8\textwidth,clip=true]{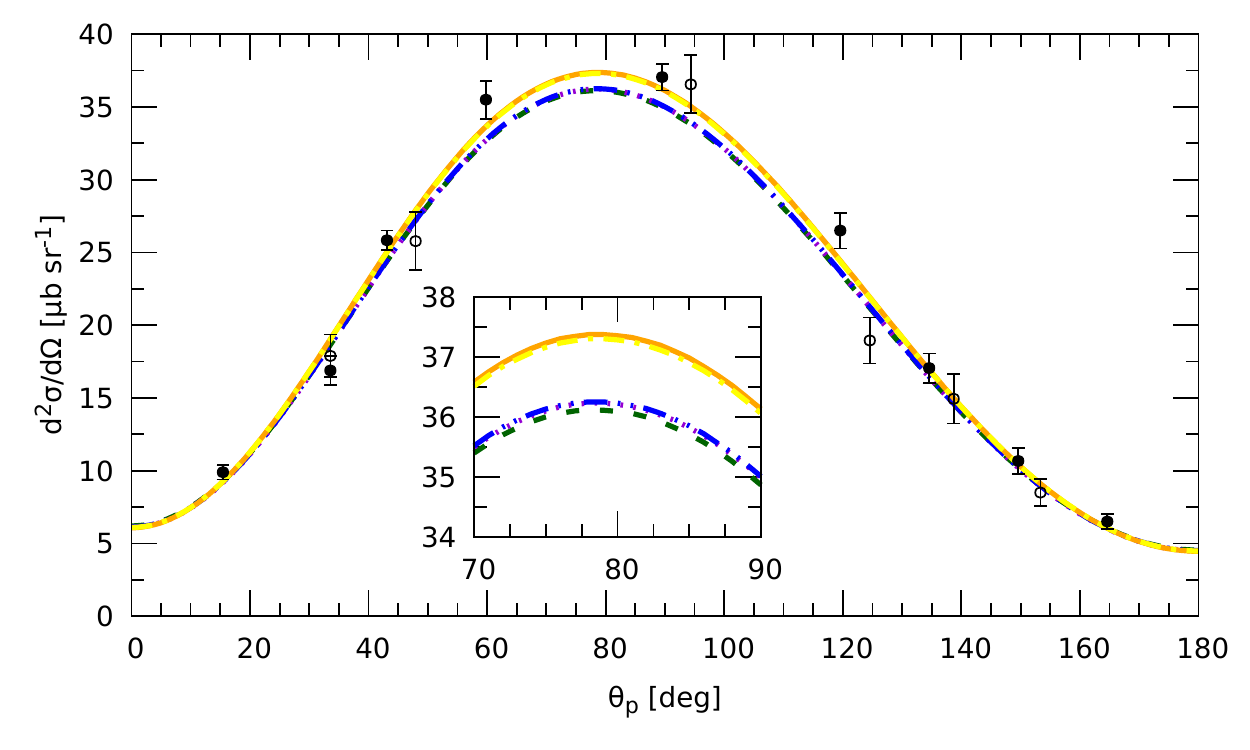}
  \end{center}
  \caption{The differential cross section $\frac{d^2\sigma}{d\Omega}$ at E$_\gamma$~=~30~MeV
	 as a function of the proton emission angle
        $\theta_p$
  calculated using the SCS (with R = 0.9~fm) and the SMS (with $\Lambda$ = 450~MeV) chiral potentials
  at higher orders of chiral expansion (N$^3$LO, N$^4$LO and N$^4$LO+ (for the SMS force only)).
  The green dashed,
  violet dotted and blue double-dot-dashed lines represent results calculated using the SMS potential
  up to N$^3$LO, N$^4$LO and N$^4$LO$^+$ respectively.
  The orange solid and the yellow dash-dotted lines are obtained using the SCS potential
  at N$^3$LO and N$^4$LO, respectively. All data points (open and filled circles)
  are taken from \cite{Ying_Experiment_Deut}.}
  \label{highord30}%
\end{figure}

\begin{figure}[h]%
  \begin{center}
  \includegraphics[width=0.8\textwidth,clip=true]{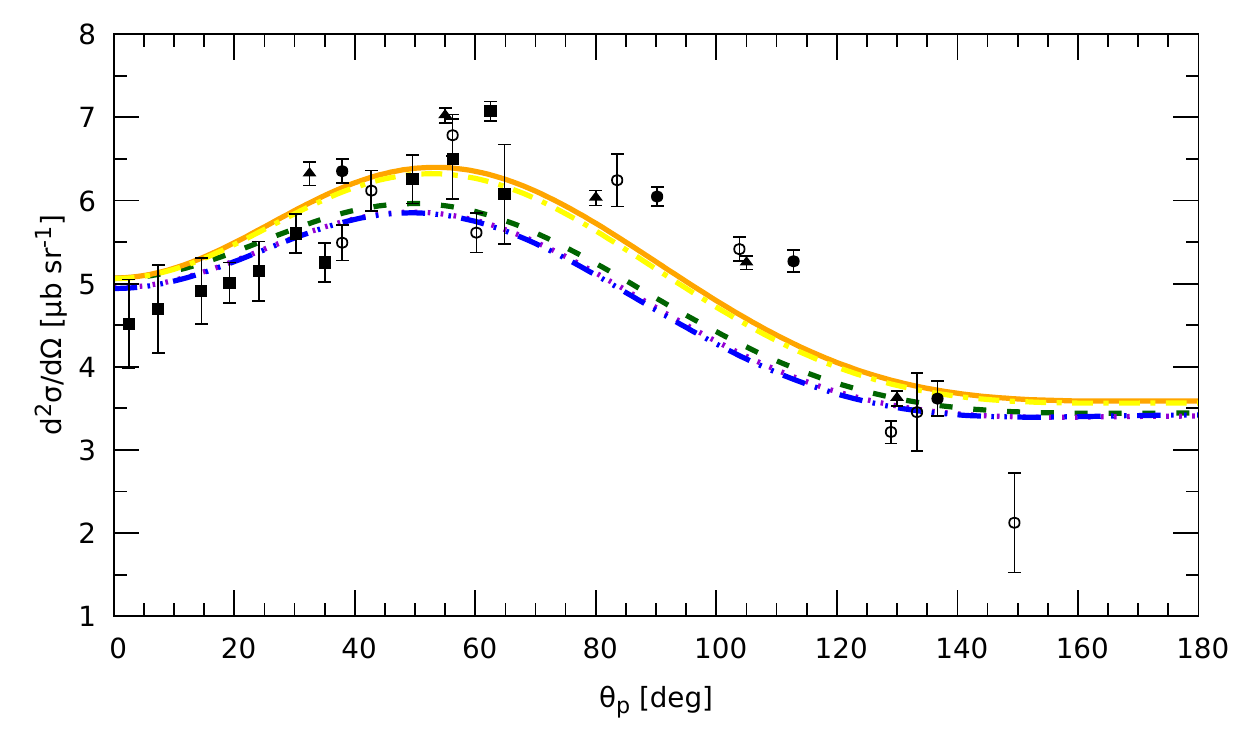}
  \end{center}
  \caption{The same as in Fig. \ref{highord30} but for E$_\gamma$~=~100~MeV.
  Data are taken from \cite{DeSanctis_Experiment_Deut} (triangles) and from
  \cite{Ying_Experiment_Deut}(open and filled circles and squares).}
  \label{highord100}%
\end{figure}

Our previous investigations \cite{skibgol2016} were devoted to the application of the older version of the
 NN potential, namely the chiral SCS force, to some electroweak processes.
 In particular, this potential applied to the deuteron photodisintegration reaction yields
 predictions for the differential cross section well converged with respect to the order of chiral
 expansion. Now we are in a position 
 to compare the outcomes from the SMS and from the SCS potentials.
 Results of our calculations of the differential cross section
 obtained using higher chiral orders (starting from N$^3$LO) of these two forces
 are presented in Figs.~\ref{highord30} and~\ref{highord100} for the photon energy 30~MeV
 and 100~MeV, respectively. It is interesting that despite the
 convergence of both potentials, curves approach different values of the cross section and
 quite a big gap between predictions of the SMS and SCS potentials is visible on both figures including 
 the inset in Fig.~\ref{highord30}. The difference between the N$^4$LO SMS and SCS cross sections 
 at $E_{\gamma}=30$~MeV reaches 1.07~$\mu$b~sr$^{-1}$ ($\approx$ 3\%) at $\theta_p=79^{\circ}$
and 0.467~$\mu$b~sr$^{-1}$ ($\approx$ 8\%) at $\theta_p=52^{\circ}$ for $E_{\gamma}=100$~MeV.
 The observed deviation can be caused by the fact that the potentials use different values of low energy constants,
 which results also in different deuteron wave functions.
 It is also possible that the lack of explicit 2N
 current contributions affects differently the predictions based on the two potentials.
Absence of such gaps for full calculations, that is ones including 
an electromagnetic current, which is complete and 
consistent with the NN interaction, will be a challenging test for 
the chiral approach.

 It is also worth mentioning that one has to be cautious about judging the agreement between 
predictions obtained with the two potentials as, beside the various regularization schemes, 
they differ in other aspects~\cite{reinkrebs2018}. Even regarding the regularization method itself 
 it is not possible to establish any one to one correspondence
 between particular values of regulators in the two spaces. 
 The prescription given in Ref.~\cite{Binder2018}, 
 $\frac{2}{R} \leftrightarrow \Lambda$, yields 
$\Lambda \approx$ 438~MeV for R = 0.9~fm.
It means that $\Lambda$ = 450~MeV and R = 0.9~fm deliver
only approximately the same regularization effect.

 Nevertheless, since for the SMS potential its version for  $\Lambda$~=~450~MeV is available, here and in other
cases
 where results based on the two potentials are compared, we use the accessible pair of regulators:
R = 0.9~fm and $\Lambda$ = 450~MeV.
 For the presented here calculations 
the comparison with experimental data cannot be used to 
judge between both forces because depending on the photon energy or the proton scattering angle either
one or the other prediction is closer to the data. Moreover, the inclusion of consistent two-body currents 
may change predictions differently for both interactions.  

\begin{figure}[h]%
  \begin{center}
    \includegraphics[width=1\textwidth,clip=true]{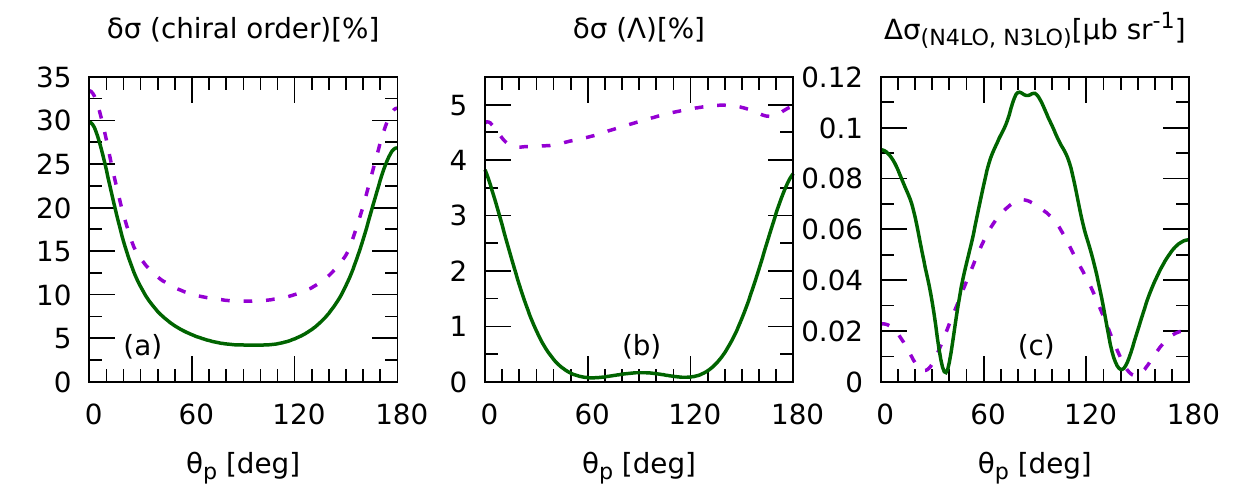}
  \end{center}
  \caption{
  The relative spread of predictions for the differential cross section 
  for the deuteron photodisintegration reaction at the initial photon energy E$_\gamma$~=~30~MeV.
  Left panel (a) represents the difference between maximum and minimum
  values of the differential cross section for all chiral orders used, divided by its average
  (from LO to N$^4$LO$^+$ for the SMS potential - green solid
  line, and from LO to N$^4$LO for the SCS  - violet dashed line). The middle panel (b) shows analogous quantity, 
  but measuring the spread with respect to the different cutoff values used 
  (from 400 MeV to 550 MeV for the SMS force and from 0.8 to 1.2 fm for the SCS potential) at fixed chiral order (N$^4$LO). 
  The right panel (c) shows absolute difference between $\frac{d^2 \sigma}{d\Omega}$
  calculated at N$^4$LO and N$^3$LO for each of two potentials.}
  \label{delta30}%
\end{figure}

\begin{figure}[h]%
  \begin{center}
\includegraphics[width=1\textwidth,clip=true]{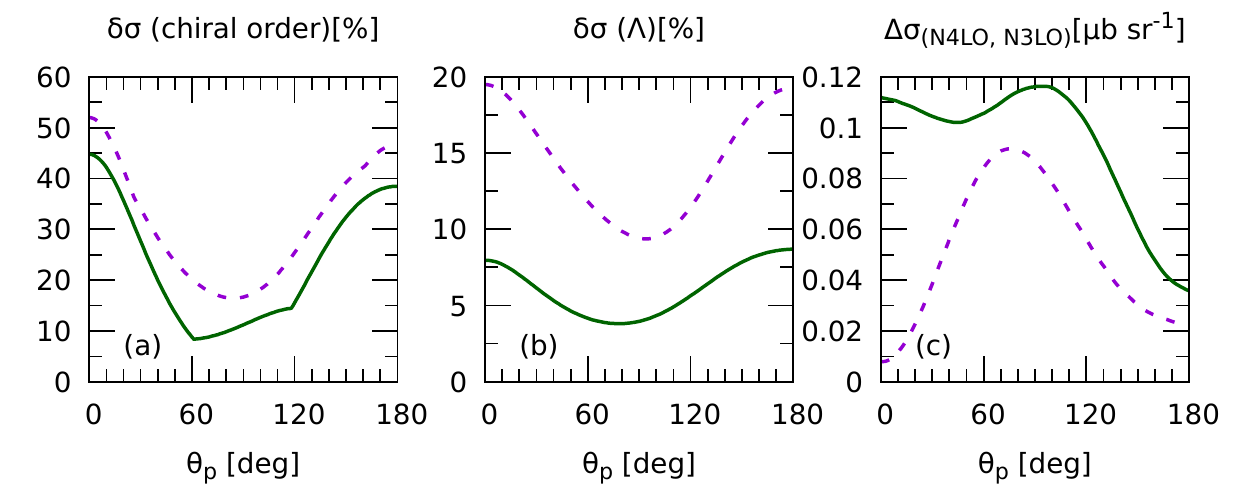}
  \end{center}
  \caption{Same as in Fig.~\ref{delta30} but for E$_\gamma$~=~100~MeV.}
  \label{delta100}%
\end{figure}

 In Figs.~\ref{delta30} and~\ref{delta100} we compare predictions based on the two
 chiral potentials, SMS and SCS, more closely. We show the relative
 difference between the differential cross section for all the presented chiral orders.
 To this end, we define $\delta\sigma({\textrm chiral\;order})$ as a difference between the maximum and minimum values of
 $\frac{d^2\sigma}{d\Omega}$ among predictions at orders from LO to N$^4$LO(N$^4$LO$^+$) for the SCS(SMS) forces
 for each angle $\theta_p$ and divide it 
by the mean value from five(six) predictions for the SCS(SMS), correspondingly.
 The resulting quantity both for the SMS (solid green line) and the SCS (dashed violet line)
 potentials 
is presented 
as a function of the proton detection angle 
in
 the left panels of Fig.~\ref{delta30} for E$_\gamma$~=~30~MeV, 
 and Fig.~\ref{delta100} for E$_\gamma$~=~100~MeV.
 In both cases the SCS result lies above the SMS one which means that the net spread with
 respect to chiral orders for the newer potential is smaller. Nevertheless this observation can also be an effect
 of the leading order predictions, which for the SMS as well as for the SCS case are far away from all the
 other results.
 Therefore it is interesting to check the absolute difference
 between differential cross sections at N$^3$LO and N$^4$LO for the two potentials, which is done
 in Figs.~\ref{delta30}c and \ref{delta100}c.
 One can see that on both plots the SMS prediction is above the SCS one for nearly all scattering angles.
 Thus the contribution from N$^4$LO is bigger for the new SMS potential.

 Figures~\ref{delta30}b and \ref{delta100}b show $\delta \sigma (\Lambda)$, 
which is an analogous quantity to this shown in Figs.~\ref{delta30}a and \ref{delta100}a, 
but now defined with respect to the values of the regularization parameter.
 That means that $\delta\sigma (\Lambda)$ is the difference between the largest and the smallest value of
 $\frac{d^2\sigma}{d\Omega}$, calculated using all values of the
 cut-off parameter ($R$ for the SCS and $\Lambda$ for the SMS) divided by its average $\sigma_{avrg(\Lambda)}$.
 From the figure it is clear that dependence on the cut-off parameter is much weaker for the new potential.
 Note that for the SCS interaction, which comprises also softer regulators, like R=1.2 fm, some
artifacts may be introduced to the potential, leading to a wider spread of the predictions.

\begin{figure}[h]%
  \begin{center}
\includegraphics[width=1\textwidth,clip=true]{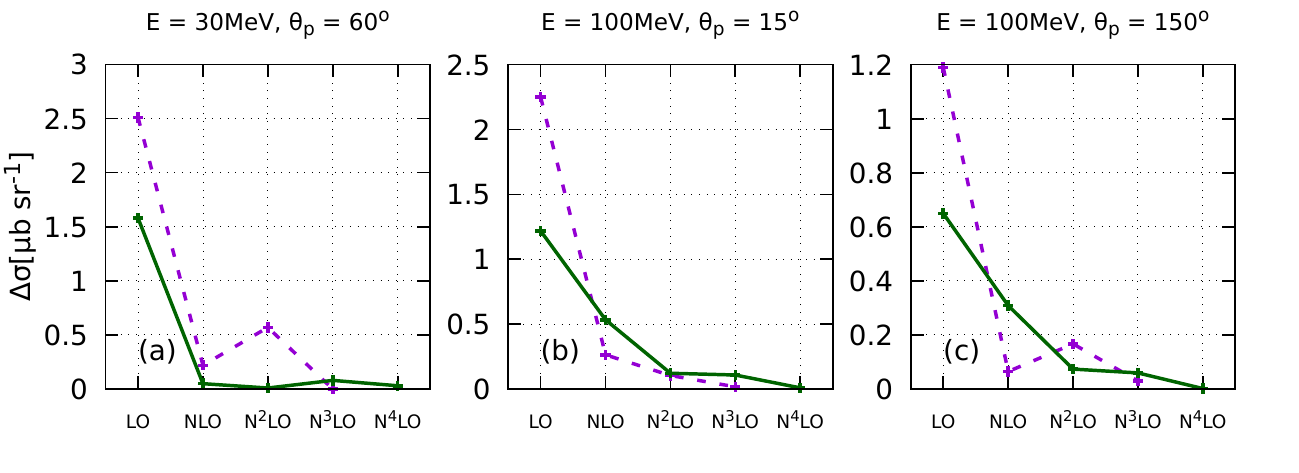}
  \end{center}
  \caption{The absolute difference between the values of the differential cross section $\frac{d^2 \sigma}{d\Omega}$ 
  taken at each two subsequent chiral orders (the one marked on the x-axis and the next one) at fixed proton 
  angle $\theta _ p$ and photon energy: $\theta _ p$ = 60$^{\circ}$ $E_\gamma$ = 30~MeV(a),
  $\theta _ p$ = 15$^{\circ}$ $E_\gamma$ = 100~MeV(b), $\theta _ p$ = 150$^{\circ}$ $E_\gamma$ = 100~MeV(c). 
  The violet dashed (the green solid) line represents results obtained using the SCS (SMS) potential
with the cutoff parameter  $R=0.9$~fm ($\Lambda=450$~MeV).
  }
  \label{diff}%
\end{figure}

 Figure~\ref{diff} demonstrates convergence of the cross section with respect to the chiral order.
 Each of the panels represents a certain combination of the photon energy and the proton emission angle:
 E$_\gamma$~=~30~MeV, $\theta_p$~=~60$^{\circ}$ at Fig.~\ref{diff}a, E$_\gamma$~=~100~MeV, $\theta_p$~=~15$^{\circ}$ at
 Fig.~\ref{diff}b, 
 and E$_\gamma$~=~100~MeV, $\theta_p$~=~150$^{\circ}$ at Fig.~\ref{diff}c. The quantity presented in this
 figure is the absolute difference between differential cross section $\frac{d^2\sigma}{d\Omega}$ 
at each two subsequent orders: the one given by the corresponding value on the $x$-axis and the subsequent one. 
For example the value with $x$-coordinate
 NLO is nothing but
 $\left|\frac{d^2\sigma}{d\Omega}|_{\scriptscriptstyle N^2LO}-
 \frac{d^2\sigma}{d\Omega}|_{\scriptscriptstyle NLO}\right|$.
We see that the SCS potential tends to converge faster, at least at presented scattering angles, since the difference 
shown in Fig.~\ref{diff}
 drops to zero earlier. This is in agreement with the results shown in Fig.~\ref{delta30}c.
Nevertheless the presence
 of an additional term N$^4$LO$^+$ in the SMS potential makes the SMS predictions converge as well, though with more
 terms included. The SMS potential does not reveal a jump between N$^2$LO and N$^3$LO predictions
as observed for the SCS force, what is caused by different off-shell behaviour of the potential.
Again, as in the case of the comparison with data, it would be interesting to see the convergence pattern for the
predictions obtained with complete currents for both potentials.

\begin{figure}[h]%
  \begin{center}
\includegraphics[width=0.8\textwidth,clip=true]{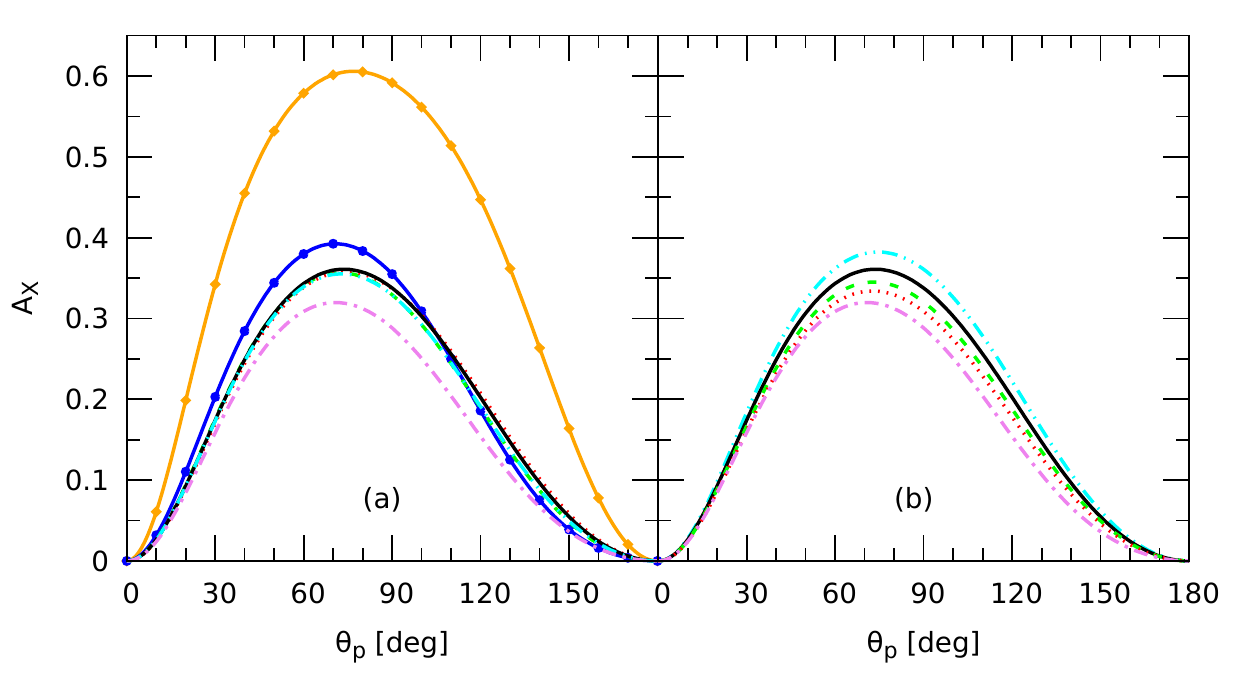}
  \end{center}
  \caption{The photon analyzing power A$_\text{X}$ as a function of the center-of-mass 
  proton detection angle 
  $\theta _ p$ for the deuteron photodisintegration process at E$_\gamma$~=~100~MeV. The left panel (a)
  shows the dependence of A$_\text{X}$ on the chiral order of the SMS potential at $\Lambda$=450~MeV.
  The right panel demonstrates the dependence of A$_\text{X}$ on the value of the cutoff parameter $\Lambda$ at N$^4$LO.
  Lines are as in Figs.~\ref{cross}a and~\ref{cross}c, respectively.}
  \label{Ax100}%
\end{figure}

\begin{figure}[h]%
  \begin{center}
\includegraphics[width=0.8\textwidth,clip=true]{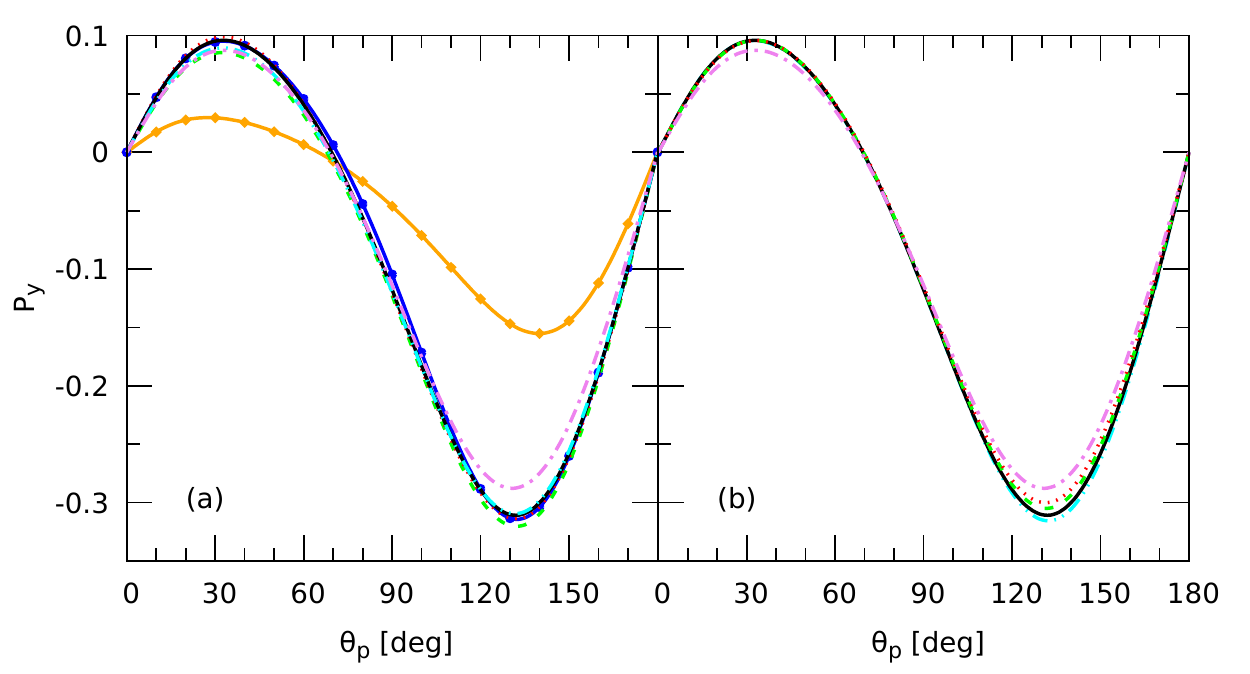}
  \end{center}
  \caption{Same as in Fig.~\ref{Ax100} but for the proton polarization P$_y$.
  }
  \label{Py100}%
\end{figure}

 In Figs.~\ref{Ax100} and \ref{Py100} we give examples of the SMS chiral force predictions
 for the polarization observables in the deuteron photodisintegration process~\cite{ArenhovelPhotodisint1991}.
 Figure~\ref{Ax100} presents the photon analyzing power A$_\text{X}$
 as a function of the outgoing proton angle $\theta_p$
 at the photon laboratory energy E$_\gamma$~=~100~MeV.
 Figure~\ref{Ax100}a shows the dependence of predictions on the order of the chiral expansion with a
 fixed regulator $\Lambda$~=~450~MeV. One can see that few lowest orders of expansion are not sufficient
 to obtain convergence of the predictions,  as only after N$^3$LO lines overlap. 
This can indicate that subsequent
orders do not bring significant contributions to the observable's final value. 
In Fig.~\ref{Ax100}b each curve corresponds to the particular value
 of $\Lambda$ (taken as 400~MeV, 450~MeV, 500~MeV, and 550~MeV) used in N$^4$LO calculations.
It is clearly visible 
that for A$_\text{X}$ the dependence on the regulator value is much
 stronger than for the maximum of the differential cross section (Fig.\ref{cross}) and amounts up to 14\%, 
 so in this case a proper choice of the $\Lambda$ parameter
 value can be important in order to obtain realistic predictions. 
 In Fig.~\ref{Py100} we present the outgoing proton polarization P$_\text{y}$ for the same reaction.
 The dependence of the predictions on the $\Lambda$ value is slightly weaker for P$_\text{y}$ than for
 A$_\text{X}$, since at the minimal values of these observables the relative difference between predictions with different
 regulator values is less than 5\% for P$_\text{y}$ (at $\theta_p = 131^{\circ}$) and reaches nearly 14$\%$ for A$_\text{X}$ 
  (at $\theta_p=72^{\circ}$).
 All our predictions
 for the deuteron photodisintegration are in a good agreement with the results obtained using
 the AV18 force and the observed differences amount to approximately 
 12\% (7\%) for A$_\text{X}$ (P$_\text{y}$) at the minima. 

\begin{figure}[h]%
  \begin{center}
\includegraphics[width=0.8\textwidth,clip=true]{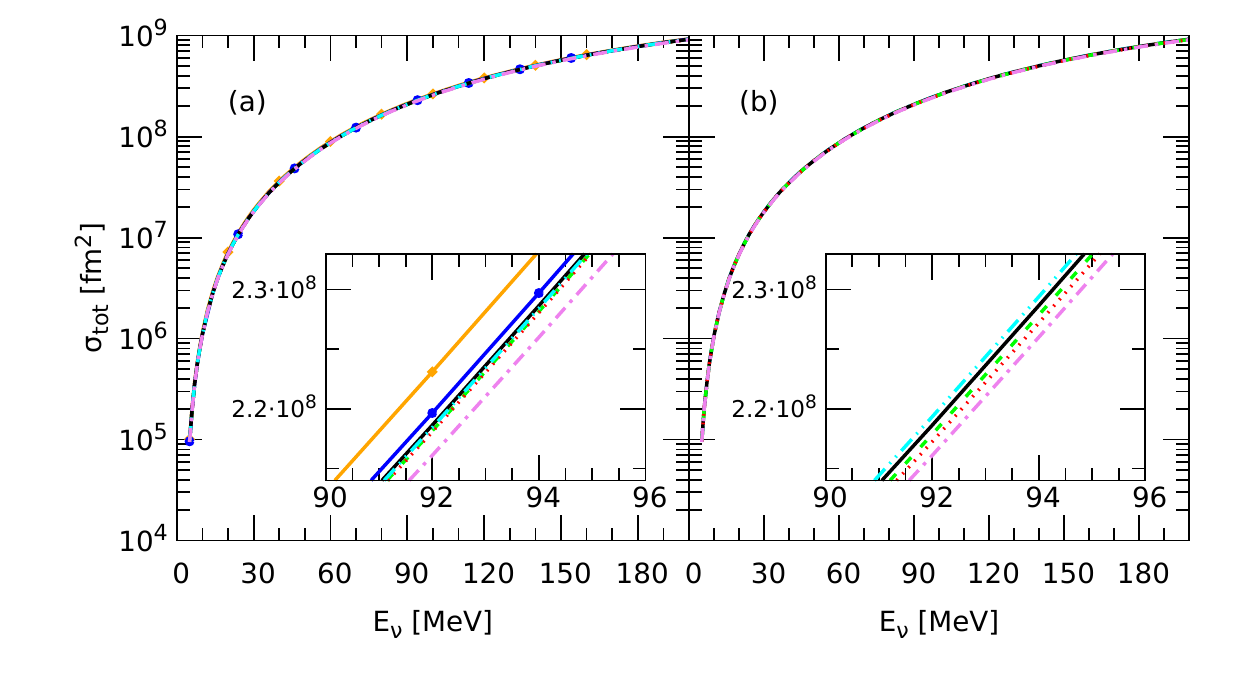}
  \end{center}
  \caption{The total cross section $\sigma_{tot}$ for the $\nu_e + ^2$H~$\rightarrow \nu_e$~+~p~+~n
  reaction as a function of the incoming neutrino energy in the laboratory system. 
 The left panel (a) presents the dependence of $\sigma_{tot}$ on the chiral order at $\Lambda$~=~450~MeV.
  Dependence of $\sigma_{tot}$ on the cutoff parameter value at N$^4$LO is presented in panel (b).
 In the left panel the orange solid line with diamonds, the blue solid line with circles, 
  the green dashed, red dotted, black solid,
  cyan double-dot-dashed and violet dot-dashed curves
  correspond to the LO, NLO, N$^2$LO, N$^3$LO, N$^4$LO, N$^4$LO+ and AV18 potential based
  predictions, respectively. In the right panel the cyan double-dot-dashed, black solid, 
  green dashed, red dotted and violet dot-dashed curves represent results with
  $\Lambda$ = 400~MeV, 450~MeV, 500~MeV, 550~MeV, and the AV18 based predictions respectively.
}
  \label{NeutrinoNC}%
\end{figure}

\begin{figure}[h]%
 \begin{center}
 \includegraphics[width=0.8\textwidth,clip=true]{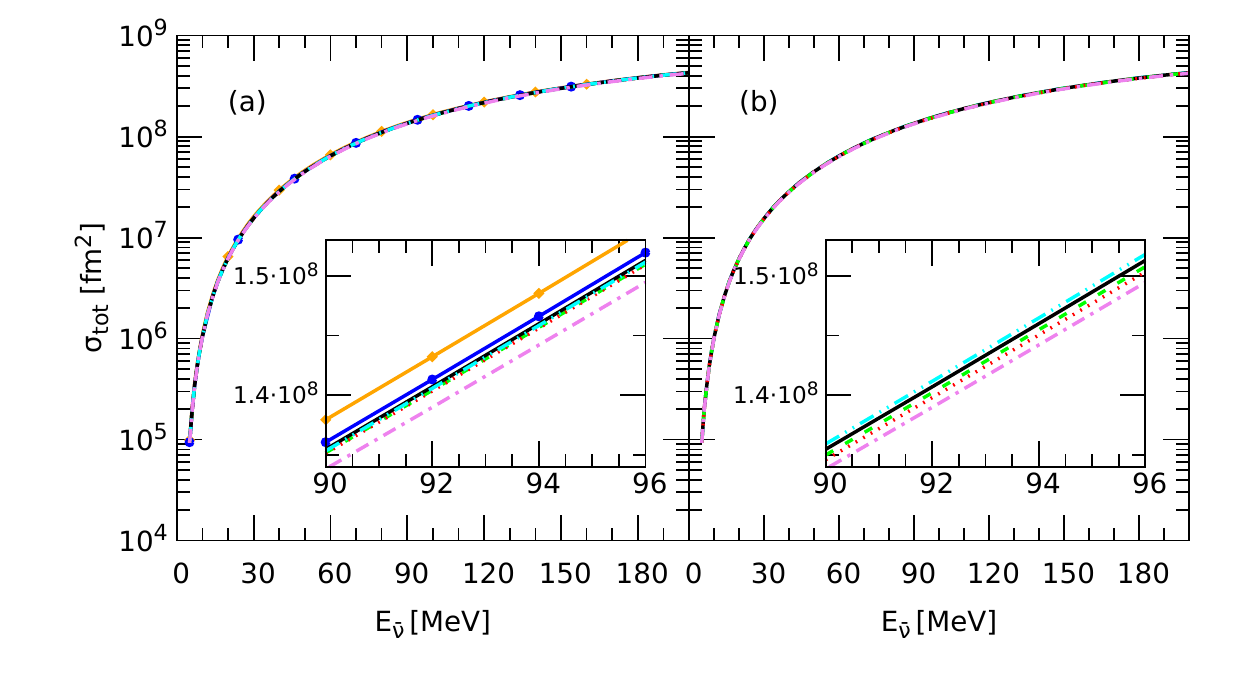}
 \end{center}
 \caption{Same as in Fig.~\ref{NeutrinoNC} but for the 
 $\bar{\nu}_e + ^2$H~$\rightarrow \bar{\nu}_e$~+~p~+~n reaction.}
 \label{AntiNeutrinoNC}%
\end{figure}

\begin{figure}[h]%
 \begin{center}
\includegraphics[width=0.8\textwidth,clip=true]{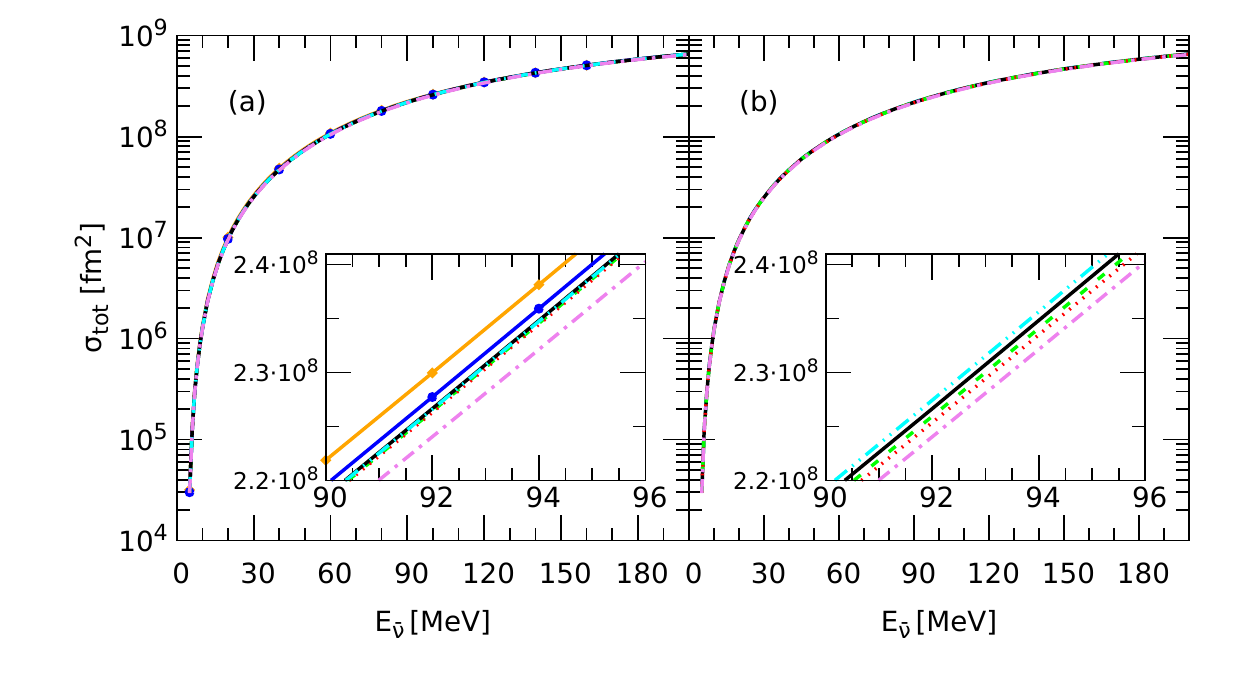}
 \end{center}
 \caption{Same as in Fig.~\ref{NeutrinoNC} but for the 
 $\bar{\nu}_e + ^2$H~$\rightarrow e^+$~+~n~+~n reaction.}
 \label{AntiNeutrinoCC}%
\end{figure}

 Now we turn to the neutrino and antineutrino induced 
 deuteron disintegration processes. For obvious reasons we restrict ourselves 
 to the total cross sections $\sigma_{tot}$. These results are obtained from the nuclear response functions
 generated on dense rectilinear grids of the $(E_{2N}, Q)$ points, where 
 $E_{2N}$ is the internal 2N energy and $Q$ is the magnitude of the three-momentum 
 transfer~\cite{Golak2019}.
 Figure~\ref{NeutrinoNC} presents predictions for the
 total cross section $\sigma_{tot}$ for the $\nu_e + ^2$H~$\rightarrow \nu_e$~+~p~+~n neutral-current 
 driven reaction.
 Again, in the left panel we show results obtained using different orders of the chiral expansion
 (from LO up to N$^4$LO$^+$, $\Lambda$=450 MeV) and in Fig.~\ref{NeutrinoNC}b the variation of the results
 with respect to the cutoff parameter value is presented. 
For the sake of comparison with the results based on a semi-phenomenological 
potential 
we also give predictions based on the AV18 interaction.
 The same dependencies but for the antineutrino induced NC disintegration $\bar{\nu}_e + ^2$H~$\rightarrow \bar{\nu}_e$~+~p~+~n
 are presented in Fig.~\ref{AntiNeutrinoNC}. The results for the charged-current induced 
 process $\bar{\nu}_e + ^2$H~$\rightarrow e^+$~+~n~+~n are demonstrated in Fig.~\ref{AntiNeutrinoCC}.

 As for our energy range $\sigma_{tot}$ takes a large spectrum of values, 
 it is hard to see the differences between curves in Figs.~\ref{NeutrinoNC}, \ref{AntiNeutrinoNC}
 and \ref{AntiNeutrinoCC} with the naked eye. 
 However, in the 
insets one can see that in the left panels of Figs.~\ref{NeutrinoNC}, \ref{AntiNeutrinoNC}
 and \ref{AntiNeutrinoCC} the two curves, representing the LO and the AV18 predictions, are separated from all the others.
 It is interesting that the relative position of the different curves 
 in the inset in Fig.~\ref{AntiNeutrinoCC}a does not remain the same trough all the energy range.
 It is so for E$_\gamma \in \left(0,145\right)$~MeV and then the LO curve swaps position with the NLO one,
 which gives the biggest predictions at the higher energies. 

 In the right panels of Figs.~\ref{NeutrinoNC}-\ref{AntiNeutrinoCC} there are results of our calculations at N$^4$LO using different regulator 
 values. To give some numerical examples: for the initial particle energy E$_{\nu \; (\bar{\nu})}$=100~MeV the relative 
differences between values of total cross section $\sigma_{tot}$
 calculated with the chiral SMS force up to the fifth order (N$^4$LO) and up to fifth order plus corrections from the sixth order (N$^4$LO$^+$) are
 0.093$\%$ for $\nu_e + ^2$H~$\rightarrow \nu_e$~+~p~+~n, 
 0.092$\%$ for $\bar{\nu}_e + ^2$H~$\rightarrow \bar{\nu}_e$~+~p~+~n and
 0.029$\%$ for the $\bar{\nu}_e + ^2$H~$\rightarrow e^+$~+~n~+~n reactions.
 The relative differences for the cross section obtained using different values of the cut-off parameter at the same
 energy, at N$^4$LO, are 0.96$\%$, 0.98$\%$ and 0.90$\%$ for the same reactions, respectively.
 It is seen that the cutoff dependence is nearly one order bigger than difference between the last two chiral orders.
Nevertheless both uncertainties remain very small which reflects low sensitivity of these inclusive observables to
 employed dynamics. The relative difference of N$^4$LO and N$^4$LO$^+$ predictions for the last reaction is approximately three times smaller than for 
 the first two in both regarded cases. Our treatment of all above mentioned weak reactions is very similar. 
 The common ingredients are the deuteron wave functions and the kinematics, which is only slightly
 modified due to the small but non-zero positron mass. The single-nucleon weak 
 neutral and charged currents are potential independent
 so they cannot explain the difference in N$^4$LO - N$^4$LO$^+$ spreads for the NC and CC driven reactions.
 However, the final states for the reactions are different.
 While in the first two reactions driven by the neutral current a neutron-proton pair emerges in the final state,
 in the third process, $\bar{\nu}_e + ^2$H~$\rightarrow e^+$~+~n~+~n, 
 a two-neutron final state is present.
 Thus the observed variation in spreads stems from the difference between 
 the neutron-neutron and neutron-proton potentials.
 From Figs. \ref{NeutrinoNC}-\ref{AntiNeutrinoCC} it is visible
 that AV18 curve here is also detached from all the other predictions.
 However the difference 
 between the predictions is small and acceptable as the potentials are constructed in quite different ways.

\section{$^3$He photodisintegration}

\begin{figure}[h]%
  \begin{center}
\includegraphics[width=1\textwidth,clip=true]{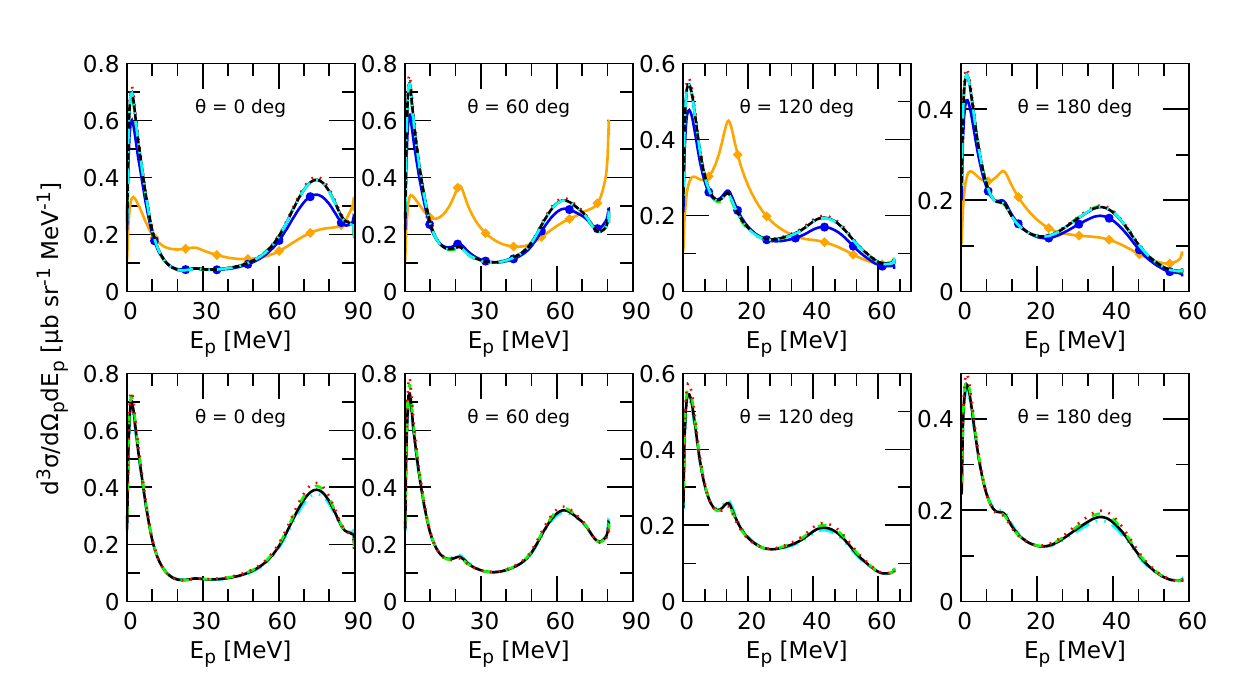}
  \end{center}
  \caption{The semi-inclusive differential cross section $\frac{d^3\sigma}{d\Omega_pdE_p}$ for the 
  $\gamma + {^3{\rm He}} \rightarrow p+p+n$ reaction at E$_\gamma$=~120~MeV as a function of the outgoing
  proton energy E$_p$ at different values of the polar angle of the outgoing proton momentum $\theta_p$.
  Top row shows the cross section
  dependence on the order of chiral expansion (at $\Lambda$~=~450~MeV), 
  the bottom row shows the dependence on the value of the cutoff parameter $\Lambda$.
  The curves are as in Fig.~\ref{cross} but the AV18 prediction is not shown here. }
  \label{CrossHe}%
\end{figure}

 In this section we discuss predictions obtained when applying the chiral force with the semi-local
 regularization in momentum space to the $^3$He photodisintegration process 
 $\gamma$~+~$^3$He~$\rightarrow$~p~+~p~+~n. As in the $^2$H case we use the Siegert theorem to go beyond the SNC approximation~\cite{Golak2005}.
In the following we neglect the three-nucleon interaction.
 The semi-inclusive differential cross section
 $\frac{d^3\sigma}{d\Omega_pdE_p}$ for the photon laboratory energy E$_\gamma$~=~120~MeV 
 is presented in Fig.~\ref{CrossHe}.
 Each of the four columns corresponds to a particular angle of the outgoing proton momentum with respect
 to the photon beam in the laboratory system  
 (0$^\text{o}$, 60$^\text{o}$, 120$^\text{o}$, and 180$^\text{o}$, respectively). Top row shows the
 dependence of the predictions on the order of chiral expansion.
 As in Figs. \ref{cross}, \ref{Ax100} and \ref{Py100} we see that it is not enough
 to take into account only leading and next-to-leading orders 
 to achieve convergence of the predictions
 and one has to include higher orders of chiral expansion.
 It is interesting to note that the older SCS potential seems to 
 converge even faster as the NLO line in Fig.~\ref{CrossHe} is farther from the higher order curves
 than one in Fig.~9 of Ref.~\cite{skibgol2016}, where the predictions
 of the SCS chiral force for the same observable are shown.
 This is similar to the already observed picture for the deuteron photodisintegration.

 The bottom row represents the cutoff dependence of the cross section which proves to be 
weak. For most of the proton energies the maximum difference
 between all predictions remains below 10$\%$. There are only exceptions for the outgoing proton angle
 0$^{\circ}$ and its energies greater than 80~MeV, where the difference 
amounts to 20$\%$ and for
 the angle 180$^{\circ}$ at the proton energies around 40~MeV, where it reaches 12$\%$.
 The cut-off dependence revealed by the SMS chiral force is weaker than it is observed for the SCS force
(compare Fig.~11 in Ref.~\cite{skibgol2016}). 

\begin{figure}[h]%
  \begin{center}
\includegraphics[width=0.8\textwidth,clip=true]{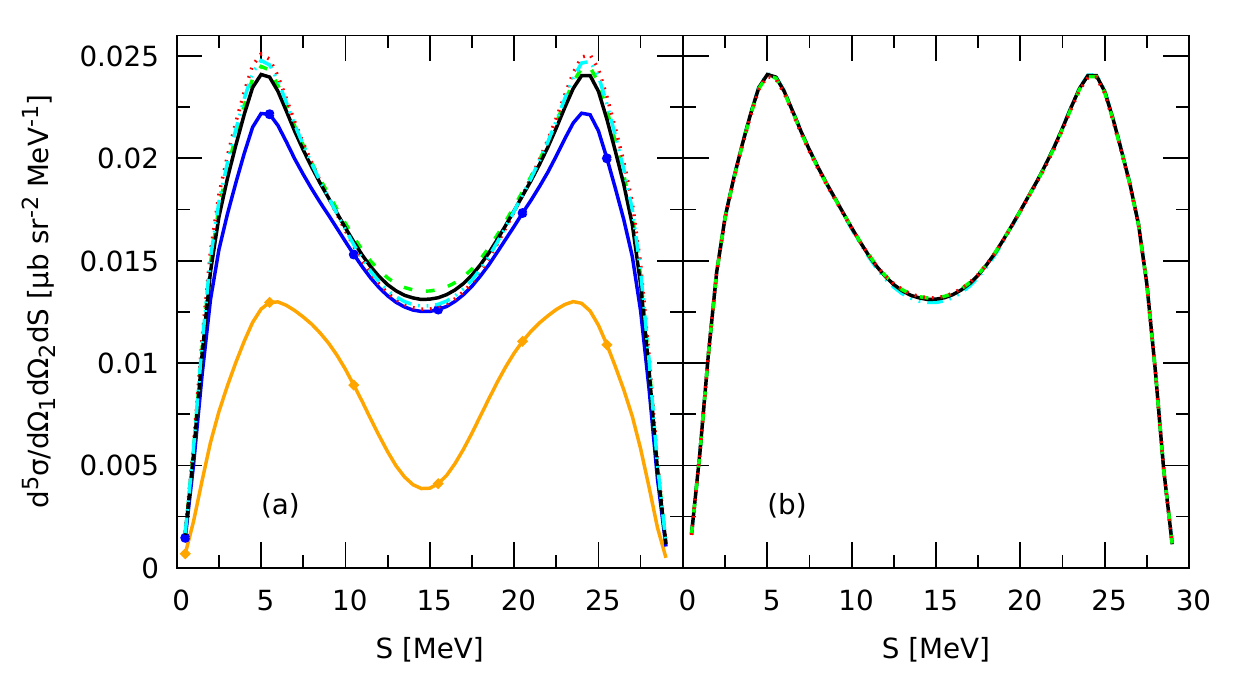}
  \end{center}
  \caption{The five-fold differential cross sections $\frac{d^5\sigma}{d\Omega_1 d\Omega_2 dS}$ for the complete kinematical configuration with the 
  two protons detected at 
  $\Theta_1$~=~15$^o$, $\Phi_1$~=~0$^o$, $\Theta_2$~=~15$^o$, $\Phi_2$~=~180$^o$ angles for
  the $^3$He photodisintegration process at the photon energy $E_\gamma$~=~40~MeV in the laboratory frame.
  The dependence on the chiral order (with $\Lambda$~=~450~MeV) is presented in the left panel (a) and 
  the results for different cutoff values (at N$^4$LO) are displayed in the right one (b).
Curves are as in Fig.\ref{CrossHe}.}
  \label{ExclHe40}%
\end{figure}

 The five-fold differential cross section $\frac{d^5\sigma}{d\Omega_1 d\Omega_2 dS}$ for the same process is presented in 
 Fig.~\ref{ExclHe40} for the photon
 laboratory energy E$_\gamma$~=~40~MeV for two protons detected at
the following polar and azimuthal angles (assuming that the momentum of 
the initial photon $\vec{p}_\gamma$ is parallel to the $z$-axis)
 $\Theta_1$~=~15$^{\circ}$, $\Phi_1$~=~0$^{\circ}$ and $\Theta_2$~=~15$^{\circ}$, $\Phi_2$~=~180$^{\circ}$.
The arc-length S of the kinematical locus in the ${\rm E}_1-{\rm E}_2$ plane,
where ${\rm E}_1$ and ${\rm E}_2$ are the kinetic energies of the two detected nucleons, is
used to uniquely define the kinematics of the three-body breakup reaction~\cite{Glockle_raport}.
 We observe that the NLO contribution is very important since it raises 
 the LO cross section
 by a factor of two. An additional shift (around 9$\%$) comes from N$^2$LO and only
 small changes are seen when N$^3$LO, N$^4$LO and N$^4$LO$^+$ force components are included.
 However, even for the highest orders (N$^4$LO and N$^4$LO$^+$) curves
 do not overlap, which suggests that full convergence is not achieved yet.
 With respect to the cut-off dependence the picture is more stable, 
 since it is very hard to distinguish individual predictions in the right panel.

\begin{figure}[h]%
  \begin{center}
\includegraphics[width=0.8\textwidth,clip=true]{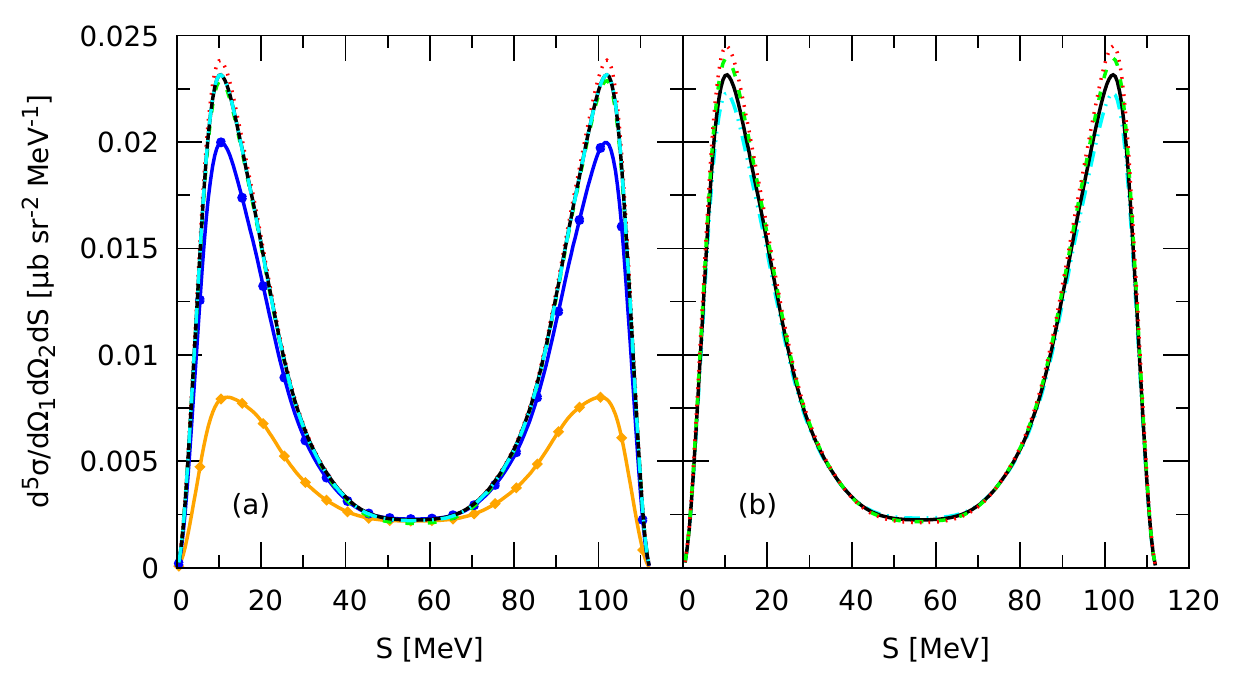}
  \end{center}
  \caption{Same as in Fig.~\ref{ExclHe40} but for $E_\gamma$~=~120~MeV.}
  \label{ExclHe120}%
\end{figure}

 Figure \ref{ExclHe120} shows the cross section for the same choice of polar and azimuthal angles
 as presented in Fig.~\ref{ExclHe40} but for the higher incoming photon
 energy E$_\gamma$~=~120~MeV. The general trends do not change with increasing energy,
 except for the fact that
 the dependence on the regularization parameter is slightly stronger which is visible at
the maxima of the cross sections around S=15~MeV and S=100~MeV. But for all other values of the arc-length 
 parameter $S$ almost full agreement between all the lines exists.

\section{Summary and conclusions}

 We presented the results of the application of the chiral NN potential 
 with the semi-local regularization
 in momentum space to a number of electroweak reactions: the $^2$H and $^3$He photodisintegrations
 and the (anti)neutrino induced deuteron breakup reactions: $\gamma + ^2{\rm H} \rightarrow {\rm p+n}$,
 $\gamma + {^3{\rm He}} \rightarrow {\rm p+p+n}$,
 $\nu_e + ^2$H~$\rightarrow \nu_e + {\rm p+n}$,
 $\bar{\nu}_e + ^2$H~$\rightarrow \bar{\nu}_e + {\rm p+n}$ and
 $\bar{\nu}_e + ^2$H~$\rightarrow e^+ + {\rm n+n}$.
 In the case of the $^3$He photodisintegration we focus on the 
properties of the applied NN potential and neglect the 
three-body force.
 
 All our results show a weaker, compared to the previous version of the Bochum-Bonn chiral potential (the SCS force),
 dependence on the cutoff parameter $\Lambda$. A small spread of results obtained with different values of $\Lambda$ 
 make the predictions based on the current interaction model more unambiguous.
 We also observe good convergence of the predictions with respect to the chiral order and the possibility 
 to include some terms from the sixth order (N$^4$LO$^+$) favorably distinguishes the SMS force from the SCS one.

 As we have no full 2N electromagnetic current consistent with the SMS interaction at our disposal,
 the Siegert theorem was used to take two-nucleon contributions in the electromagnetic
 current operator at least partly into account for the photodisintegration processes. 
 As a consequence, the incomplete electromagnetic current operator leads to
 some problems with the data description, as seen in Fig.~\ref{cross}.
 We expect that future application of the electromagnetic current operator fully consistent 
with the 2N potential will significantly improve the agreement with the data.

 The presented polarization observables for the deuteron photodisintegration processes ($A_X$ and $P_y$) are the ones, where the 
 slowest convergence with respect to the chiral order and the strongest dependence on the regularization
 parameter is noticed. For instance for the deuteron analyzing powers ($T_{11}$, $T_{20}$, $T_{21}$ and $T_{22}$)
 the convergence is very fast (above the leading order) and the cutoff variation is negligible.
 
 The same picture is also valid for the investigated here total cross sections for the weak $^2$H disintegrations via
neutral or charged currents: the convergence with respect to the chiral order is quite rapid and the cutoff dependence 
is weak.

 The new SMS potential has a number of practical advantages in comparison to the older chiral forces. Its predictions show
 weaker dependence on the cut-off parameter and good convergence with respect to the order of chiral NN potential.
 There is still room for improvement in our calculations. The main drawback of the present formalism
 is the lack of the explicit electroweak current operator entirely consistent with the 2N SMS interaction 
 as well as the omission of the three-nucleon force for 
 the $^3$He disintegration. 
In addition, the future complete studies should be supplemented with an
analysis of truncation errors using the Bayesian approach, which
would allow one to draw more reliable conclusions
about the convergence pattern of chiral EFT predictions for these reactions.
Nevertheless, our results with the simplified Hamiltonian reveal the usefulness of the SMS chiral force
for studies of various electroweak processes in few-nucleon systems in the near future. 

\section{Acknowledgements} 
This work is a part of the LENPIC project and was supported by the Polish National Science Center under Grants No. 2016/22/M/ST2/00173 
and 2016/21/D/ST2/01120. 
It was also supported in part by BMBF (Grant No. 05P18PCFP1) and by DFG through funds provided to the Sino-German CRC 110 “Symmetries and the Emergence
of Structure in QCD” (Grant No. TRR110).
The numerical calculations were partially performed on the supercomputer cluster of the JSC,
J\"ulich, Germany.


\bibliography{mybib}


\end{document}